\documentclass[epj,final,twocolumn]{svjour}

\usepackage{graphicx}
\usepackage{bm}

\begin{document}

\title{$\delta$-pairing forces and collective pairing vibrations}
\author{Kamila Sieja\thanks{e-mail: ksieja@hektor.umcs.lublin.pl}, 
 Andrzej Baran,  
 and Krzysztof Pomorski
 }
\institute{Institute of Physics, Maria Curie-Sk{\l}odowska University\\
ul. Radziszewskiego 10, 20-032 Lublin, Poland}
\date{Received: April 17, 2003 / Revised version: September 8, 2003 } 

\abstract{
The collective pairing hamiltonian is obtained
in the framework of the generator coordinate method in the gaussian overlap approximation
with a slightly modified BCS function used as a generator function. 
The collective variable $\alpha$, measuring the monopole moment of the pairing field,
and the gauge transformation angle $\phi$ are chosen as generator coordinates. 
The vibrational ground states are calculated by diagonalisation
of the collective pairing hamiltonian in the harmonic oscillator basis.
} 

\maketitle

\section{Introduction \label{int}}

The collective pairing hamiltonian was originally introduced in Ref.
\cite{Bes}, where the intrinsic deformation of the pairing field  $\alpha$,
related to the BCS gap parameter $\Delta$, and the gauge transformation angle
$\phi$ were used as collective variables. The pairing hamiltonian was then
derived in framework of the cranking  approximation.  A more general method
which determines  wave functions and the collective  hamiltonian is the Generator
Coordinate Method (GCM).  The GCM derivation of the collective hamiltonian with
the BCS functions as the generator functions were already  performed in Ref.
\cite{pom1,pom2,pom3}, where the Gaussian Overlap Approximation (GOA) of the
generator wave functions was used.  The collective pairing Schr\"odinger
equation was constructed using the single-particle plus  monopole pairing
hamiltonian.

The aim of this paper is to derive collective vibrations of the pairing field
in GCM+GOA approach for the hamiltonian based on the single-particle Nilsson 
potential \cite{nil} and the $\delta$-pairing interaction
\cite{quentin,baran,meng}. The latter requires introducing a suitable collective
variable different from the monopole pairing gap used in \cite{pom1}.  Here we
have used  the same generator coordinate $\alpha=\sum u_k v_k$ as in Ref.
\cite{Bes}, which as suggested in Ref. \cite{meyer} is a natural choice  for
this type of collective motion.


The paper is organized as follows: In Sect. \ref{roz1} we discuss the
properties of the $\delta$-pairing interaction. In Sect. \ref{roz2} the
collective coordinates and the form of the collective pairing  hamiltonian are
introduced. The formulae for the GCM+GOA mass parameters are compared to the
cranking ones. The method of diagonalising the collective pairing
hamiltonian is described in Sect. \ref{roz3}. Sect. \ref{roz4} contains
numerical results  and Sect. 6 conclusions. In Appendix \ref{sec-appendix}
we discuss the multipole expansion of the  $\delta$-force and demonstrate
validity of our choice of $\alpha$ as the collective coordinate.

\section{$\delta$-pairing forces}\label{roz1}
The nuclear mean-field hamiltonian with the residual pairing interaction
can be written as
\begin{equation}
\label{tmb}
\hat H= \hat H_{s.p.}+\hat H_{pair},
\end{equation}
where
\begin{equation}
\hat H_{s.p.} = \sum_{k>0}\langle k|\hat h|k\rangle(c_k^\dagger c_k+
                c_{\bar k}^\dagger c_{\bar k}),
\end{equation}
\begin{equation}\label{ph}
\hat H_{pair} = -\sum_{k,l>0}V_{k \bar k l \bar l}c_k^\dagger 
                c_{\bar k}^\dagger c_{\bar l}c_l\,.
\end{equation}
The summation runs over the eigenstates $|k\rangle$  of the single-particle 
hamiltonian $\hat h$. The antisymmetrized matrix element of the pairing
interaction  $V_{k \bar k l \bar l}$ in the pairing hamiltonian (\ref{ph}) is
given by the following expression
\begin{eqnarray}
V_{k\bar k\bar l l} = 
\quad \int d^3 r_1d^3 r_2 \sum_{\sigma_1\sigma_2}\Phi_k^*(\vec r_1,\vec\sigma_1)
\quad \Phi_{\bar k}^*(\vec r_2,\vec\sigma_2)  \nonumber \\
V^{\tau}(\vec r_1,\vec\sigma_1;\vec r_2,\vec\sigma_2) 
\nonumber\\
\left[\Phi_l(\vec r_1,\vec\sigma_1)\Phi_{\bar l}(\vec r_2,\vec\sigma_2)-
\Phi_{\bar l}(\vec r_1\vec\sigma_1)\Phi_l(\vec r_2,\vec\sigma_2)\right]\,.
\end{eqnarray}   
 Here $\Phi_k(\vec r,\vec\sigma)$ is the single-particle eigenfunction of 
 $\hat h$ in the space $(\vec r)$ and spin $(\vec\sigma)$ representation
and $\Phi_{\bar k}(\vec r,\vec\sigma)$ is its time reversal counterpart.
$V^{\tau}(\vec r_1,\vec\sigma_1;\vec r_2,\vec\sigma_2)$
is the $\delta$- pairing force \cite{quentin}
\begin{equation}
V^{\tau}(\vec r_1,\vec\sigma_1;\vec r_2,\vec\sigma_2)=V_0^\tau  
\frac{1-\vec\sigma_1\cdot\vec\sigma_2}{4}\delta(\vec r_1-\vec r_2)\,.
\end{equation}
In the following we consider the proton-proton (p-p) and neutron-neutron (n-n) 
part of the interaction only. Taking this into account one obtains:
\begin{equation}
G_{kl}=V_{k\bar kl\bar l}=V_0
\int d^3r \rho_k(\vec r)\rho_l(\vec r)\,,
\end{equation}
where
\begin{equation}
\rho_k(\vec r) = \left| \Phi_k(\vec r)\right|^2\,
\end{equation}
and $V_0$ is the pairing interaction strength as adjusted in Ref. \cite{sieja}. 
 The pairing gap equations become
\begin{equation}
\Delta_k=\frac{1}{2}\sum_{l>0} G_{kl}\frac{\Delta_l}
         {\sqrt{(e_l-\lambda)^2+{\Delta_l}^2}}\,,
\label{delta}                                                  
\end{equation}
where $e_l$ is the single particle energy and the Fermi
level $\lambda$ is determined from the particle number equation
\begin{equation}
 N = \sum_{k>0}\left ( 1 - \frac{e_k-\lambda}{\sqrt{(e_k-\lambda)^2+
    {\Delta_k}^2}} \right)\,.                               
\label{n} 
\end{equation}
The coupled set of Eqs. (\ref{delta}) and (\ref{n}) is solved numerically
by the appropriate iteration procedure.
 
\section{Collective pairing hamiltonian}\label{roz2}
Following Ref. \cite{bohr} we construct  the monopole pairing operator 
(see Appendix) which characterizes the amplitude of excitations
connected with vibrations of the pair density
\begin{equation}
\label{mm}
\hat A = \frac{1}{2}\sum_{k>0}\left( e^{-2i\phi}c_k^\dagger c_{\bar k}^\dagger + 
e^{2i\phi}c_{\bar k}c_k \right)\,.
\label{eq-totham}
\end{equation}
The pair condensate can be described by the mean value of the operator 
(\ref{mm}) in the BCS like state
\begin{equation}
\label{BCS}
|\alpha\phi\rangle = e^{iN\phi}\prod_{k>0}
\left(u_k+v_k e^{-2i\phi}c_k^\dagger c_{\bar k}^\dagger 
\right)|0\rangle\,,
\label{bcsst} 
\end{equation}
where $N$ is the number of particles, $\phi$ is the gauge angle and
the average pairing gap is given by the expectation value of the operator 
(\ref{mm})
\begin{equation}
\alpha = \langle\alpha\phi|\hat A|\alpha\phi\rangle = \sum_{k>0} u_k v_k\,.
\end{equation}
Using (\ref{BCS}) as a generator function and $\alpha$, $\phi$ as generator 
coordinates and following the steps of Ref. \cite{pom1} the collective 
hamiltonian
   \begin{eqnarray}
\hat\mathcal{H}_{coll} &=& -\frac{\hbar^2}{2\sqrt {{\rm det}
\gamma_{\alpha\alpha}}}
\frac{\partial}{\partial\alpha}\sqrt{{\rm det}\gamma_{\alpha\alpha}}
\mathcal{M}_{\alpha\alpha}^{-1}
\frac{\partial}{\partial\alpha} \nonumber \\
&& \quad-\frac{1}{2}\hbar^2\mathcal{M}_{\phi\phi}^{-1}\frac{\partial^2}
{\partial\phi^2} \nonumber  \\
&& \quad-i\hbar\frac{{\rm Im}\langle\alpha\phi 
|\frac{\stackrel\leftarrow\partial}{\partial\phi}\hat H|
\alpha\phi\rangle}{\gamma_{\phi\phi}}\frac{\partial}{\partial\phi}\nonumber \\
&& \quad +V(\alpha)\,,
\label{hcoll}
\end{eqnarray}
is derived. In Eq. (\ref{hcoll}) $\gamma_{\alpha\alpha}$, $\gamma_{\phi\phi}$
are related to the widths of the Gaussian overlap and
$\mathcal{M}_{\alpha\alpha}^{-1}$, $\mathcal{M}_{\phi\phi}^{-1}$ are the
components of the inverse mass tensor.  The quantities appearing in Eq.
(\ref{hcoll}) are analogous to the general  expressions obtained for the
$\Delta$ and $\phi$ coordinates in the  monopole pairing case \cite{pom1}.
$V(\alpha)$ is the collective pairing potential equal to
\begin{equation}
V(\alpha) = \langle\alpha\phi|\hat H|\alpha\phi\rangle-{\cal E}_0\,,
\label{eqv}
\end{equation}
where ${\cal E}_0$ is the so-called zero-point energy.
The collective pairing hamiltonian (\ref{hcoll}) is hermitian with 
the Jacobian
\begin{equation}
d\tau = \sqrt{\gamma_{\alpha\alpha}\gamma_{\phi\phi}}\,\alpha d\alpha\, d\phi\,.
\end{equation}
It means that the eigenfunctions of (\ref{hcoll}) should be orthogonal 
with the above measure.    

The mean value of the hamiltonian (\ref{tmb})
is evaluated using the constraints $N={\rm const}$ and $\alpha={\rm const}$ 
which leads to minimization
of the average value of the operator
\begin{equation}
\hat H'=\hat H-\lambda(\hat N-\langle\alpha\phi|\hat N|\alpha\phi\rangle)
-\xi(\hat A-\langle\alpha\phi|\hat A|\alpha\phi\rangle)\,, 
\end{equation} 
calculated in the BCS state (\ref{bcsst}).
Here $\lambda$ and $\xi$ are Lagrange multipliers.

The expectation value of the BCS hamiltonian in Eq. (\ref{eqv}) is given by
\begin{eqnarray}
E_{BCS} &=& \langle \alpha\phi | \hat H | \alpha\phi \rangle =\nonumber \\
        &=& 2\sum_{k>0}e_k v_k^2-\frac{1}{2}\sum_{k>0}
        \frac{\Delta_k^2}{E_k}-\sum_{k>0} G_{kk}v_k^4\,,
\label{eqbcs}
\end{eqnarray}
where $E_k=\sqrt{(e_k-\lambda)^2+\Delta_k^2}$ is the quasiparticle energy. 

The final expressions for the nonvanishing components of the metric and mass 
tensors are
\begin{equation}
\gamma_{\alpha\alpha} = \sigma^2\sum_k\frac{(e_k-\lambda)^2}{16E_k^4}\,,
\end{equation} 
\begin{equation}
\gamma_{\phi\phi} = \sum_k\frac{\Delta_k^2}{E_k^2}\,,
\end{equation}
\begin{equation}
\mathcal{M}_{\alpha\alpha}^{-1} = 
\sum_k  E_k \sigma^2\left/\right.\gamma_{\alpha\alpha}^2\,,
\label{maa}
\end{equation}
\begin{equation}
\mathcal{M}_{\phi\phi}^{-1} = 
\sum_k\frac{\Delta_k^2}{E_k}\left/\right.\gamma_{\phi\phi}^2\,,
\end{equation}
where $\sigma$ is equal to 
\begin{equation}
\sigma = \left(\sum_k \frac{(e_k-\lambda)^2}{4E_k^3}\right)^{-1}\,.
\end{equation}
The zero-point energy appearing in Eq. (\ref{eqv}) reads
\begin{eqnarray}
{\cal E}_0 &=&
{\cal E}_0^{\alpha}+{\cal E}_0^{\phi} \nonumber\\
&=& \frac{1}{2}\left({\sum_k\frac{(e_k-\lambda)^2}{16E_k^3}\sigma^2}
\left/\right.\gamma_{\alpha\alpha}+\sum_k\frac{\Delta_k^2}{E_k}\left/\right.
\gamma_{\phi\phi}\right).
\label{e0}
\end{eqnarray}
The first term in Eq. (\ref{e0}) represents the collective correlations  
in the ground state energy whereas the second corresponds to the approximate 
particle number projection \cite{pom3}.

An alternative method of treating collective vibrations is the cranking 
model \cite{bel,wilets}. The mass parameters in this case read 
\begin{equation}
\mathcal{M}_{\alpha\alpha}^{cr} = 2\sigma^2\sum_k\left(\frac{e_k-\lambda}{16E_k^3}\right)^2
\label{bcr}\,,
\end{equation}
\begin{equation}
\mathcal{M}_{\phi\phi}^{cr} = \sum_k\frac{\Delta_k^2}{E_k^3}\,
\end{equation}
and the collective potential is usually assumed to be equal to $E_{BCS}$ 
(\ref{eqbcs}).

\section{Diagonalisation of the collective pairing hamiltonian}\label{roz3}

Having found the collective potential, the mass and the metric tensors, we can 
evaluate the collective hamiltonian numerically. Only the vibrational spectra 
will be constructed here as the quasirotational states correspond
to the bands built from the ground states function of the
neighbouring even- even isotopes (or isotones).
  In the first step we perform a transformation from the $\alpha$ coordinate
to a new variable $x$ in which the mass parameter $\mathcal{M}_{xx}$ is nearly constant
\begin{equation}
\mathcal{M}_{xx} \approx
\mathcal{M}_{\alpha\alpha}\left(\frac{\partial\alpha}{\partial x}\right)^2
=\rm{const}\,.
\end{equation}
The mass parameter $\mathcal{M}_{\alpha\alpha}$ is a rapidly decreasing function
of $\alpha$ and can be approximated by the function
\begin{equation}
\mathcal{M}_{\alpha\alpha}=\frac{b}{(\alpha+\alpha_0)^2}\,,
\end{equation}
where $\alpha_0$ and $b$ parameters are chosen to approximate the mass parameter in
the best way. Variables $x$ and $\alpha$ are connected through the equation
\begin{equation}
x = \sqrt{b/\mu}\,{\rm ln}\left(1+\frac{\alpha}{\alpha_0}\right)\,,
\end{equation}
where $\mu$ is an arbitrary constant.
The basis states used to diagonalise the hamiltonian (\ref{hcoll})
 are generated by the harmonic oscillator hamiltonian
\begin{equation}
\hat H_B = -\frac{\hbar^2}{2\mathcal{M}_{xx}}\frac{d^2}{dx^2}+\frac{1}{2}
\mathcal{M}_{xx}\omega^2x^2\,,
\end{equation}
with frequency $\omega$ determined from the plateau condition of the energy 
$\partial E/\partial\omega\approx0$, where $E$ is the ground state energy
of the collective hamiltonian (\ref{hcoll}).
In our case only the even eigenstates are picked (see \cite{pom1})
\begin{equation}
\hat H_B\psi_i = (i+\frac{1}{2})\psi_i\,,\quad i=0,2,4,\dots\,.
\end{equation}
These eigenstates which satisfy the usual normalization conditions
\begin{equation}
\int_0^{\infty} \psi_i\psi_j dx=\delta_{ij}\,
\end{equation}
are employed to calculate the matrix elements of the collective hamiltonian
(Eq. \ref{hcoll}).
The digonalization was performed in a basis of $N_{max}=19$ oscillator
shells.

\begin{center}
\begin{figure}
\includegraphics[scale=0.8]{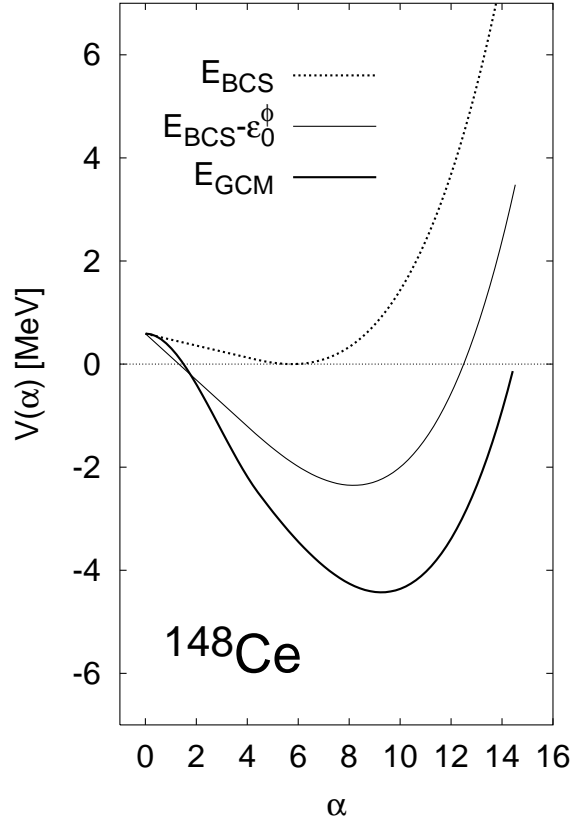}
\caption{
Pairing potentials: the BCS energy (dashed line), the GCM collective potential 
(solid line) and the particle number corrected BCS energy 
$E_{BCS}-{\cal E}_0^\phi$ (thin solid line) as functions of the collective 
coordinate $\alpha$ for protons in $^{148}$Ce. Zero value on the abscissa 
corresponds to the minimum of the BCS energy.}
\end{figure}
\end{center}

\begin{center}
\begin{figure}
\includegraphics[scale=0.8]{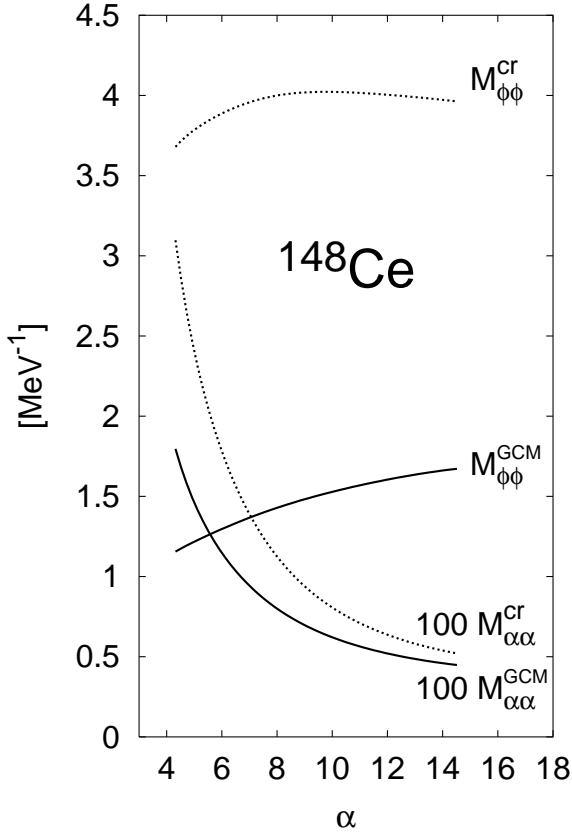}
\caption{
Components of GCM and the cranking mass parameters for protons in $^{148}$Ce.
The $M_{\alpha\alpha}$ components are multiplied by the factor 100. 
The components  
$M_{\phi\phi}$ are in natural units.
}
\end{figure}
\end{center}

\section{Results}\label{roz4}

We have considered a few nuclei in the rare earth region. The ${\rm A}=165$ 
parameter set of the single particle Nilsson potential is used \cite{nil1}. 
The $\delta$-pairing strengths as obtained in Ref. \cite{sieja} are
\begin{eqnarray}
V_0^{\rm neutrons} &= 230\, {\rm MeV\, fm^3}\,,\nonumber\\
V_0^{\rm protons}  &= 240\, {\rm MeV \, fm^3}\,.
\label{vo}
\end{eqnarray}
The fit of the $V_0$ parameters was done for the pairing window with
$2\sqrt{15\,n}$ levels closest to the Fermi energy where
$n=N$ for neutrons and $n=Z$ for protons \cite{nil1}.

Figure 1 shows the pairing potentials for BCS and GCM models.
The dashed line represents the pure BCS energy (\ref{eqbcs}) as a function of the 
collective coordinate $\alpha$ whereas the solid line corresponds to
the GCM collective potential (\ref{eqv}). The particle number corrected BCS 
energy, {\it i.e.}, the one with extracted ${\cal E}_0^\phi$ energy
is also shown (thin solid line). Zero value on the abscissa corresponds to the 
minimum of the BCS energy. The particle number projected energy is about $2$ MeV deeper
than the BCS energy minimum and its position is shifted in the direction 
of larger $\alpha$.
Analogously, the GCM ground state energy which additionally contains the zero point energy 
corresponding to the pure $\alpha$ vibrations has a minimum shifted by a similar amount 
towards larger $\alpha$.

The $M_{\alpha\alpha}$ and $M_{\phi\phi}$ components of the mass tensor 
for both GCM and the cranking model are shown in Fig. 2.
The ratio of the $\alpha$ component
of the cranking mass (dashed line) to the GCM mass (solid line) is close to $2/3$ 
in the vicinity of the minimum of the GCM potential. The decrease of the 
collective mass with $\alpha$ (or $\Delta$) has a significant influence on
the spontaneous fission half lives of heavy nuclei, as shown for $\Delta$ collective coordinate, 
as well as on the height of the fission barriers (see {\it e.g.}, \cite{stasz}). 
Similarly in case of the collective Bohr model calculations \cite{zajac1,zajac2}
the inclusion of coupling of quadrupole vibrations improves significantly the agreement
with the experimental data.

\begin{center}
\begin{figure}
\includegraphics[scale=0.8]{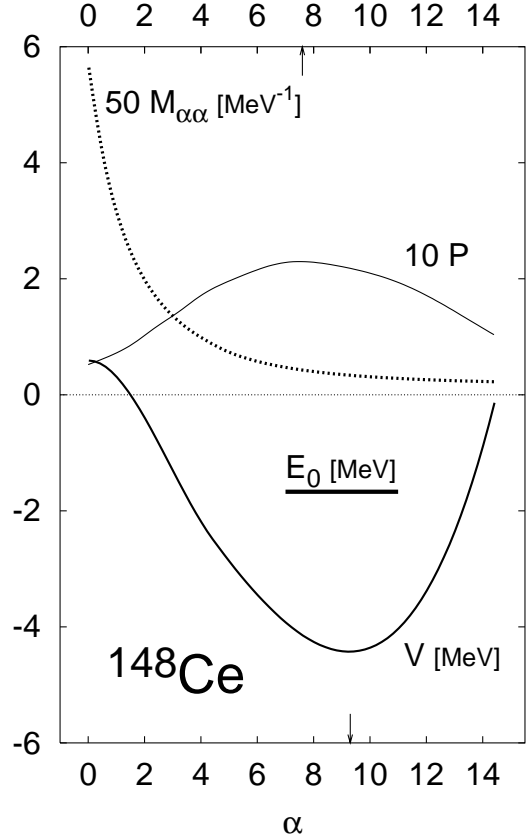}
\caption{
The collective potential $V(\alpha)$ (solid line), the GCM mass parameter 
$M_{\alpha\alpha}(\alpha)$
(dotted line) and the probability density 
$P=|\Phi_0(\alpha)|^2\sqrt{\gamma_{\alpha\alpha}\gamma_{\phi\phi}}$ (thin solid line) 
of the ground state for protons in $^{148}$Ce. The ground state energy $E_0$ is marked.
The arrows indicate the positions of the minimum of the potential and the maximal value 
of the probability distribution on the $\alpha$ axis.}
\end{figure}
\end{center}               
In Fig. 3 we show the collective potential $V(\alpha)$ (solid line), the GCM mass parameter $M_{\alpha\alpha}(\alpha)$
(dotted line) and the probability density $P=|\Phi_0(\alpha)|^2\sqrt{\gamma_{\alpha\alpha}\gamma_{\phi\phi}}$ 
of the ground state (thin solid line). 
The short line segment in the middle of the figure 
marks the position of the ground state energy. The arrows indicate the positions of the minimum of the potential 
(bottom scale)
and the maximal value of the probability distribution (top scale).
 The $\alpha$ value of the equilibrium is 9.3 and the most 
probable value is 7.6.  The latter is shifted  towards smaller
$\alpha$ values that implies the increase of the
$M_{\alpha\alpha}$ by $1.5$ factor on average.
This behaviour is common for the majority of nuclei in the considered region. 
As shown in Ref. \cite{zajac1} in case of monopole pairing the use of the 
most probable value
of the gap parameter (instead of the BCS gap) leads to a considerable improvement of
the predictive power of the Bohr hamiltonian.

\section{Conclusions}
The first excited collective pairing vibrational states for even-even 
nuclei in rare earth region
appear at energies close to 2.5 MeV for protons and 4.5 MeV for neutrons.
Usually they have higher
energies than two quasiparticle excitations and consequently one has to 
include these correlations in the ground state, only.

Since the pairing vibrations are strongly coupled
with shape degrees of freedom of the nucleus
it is hard to compare the results (which are similar to that obtained with the coordinate $\Delta$)
 with experimental data. 

Finally,  it should be emphasized that the role of the collective coordinate
 $\alpha$ is analogous to the $\Delta$ coordinate considered earlier. 

\begin{acknowledgement}
We are indebted to Jerzy Matyjasek and Paul Stevenson for careful reading
 of the manuscript and valuable comments. 
\end{acknowledgement}

\appendix

\section{Multipole expansion of $\delta(\bm r)$-force\label{sec-appendix}}

In the following we consider the short range interaction operating between 
equivalent nucleons ($T=1$, protons or neutrons)
\begin{eqnarray}
V_{12}(\delta)&=&V(\vec r_1, \vec\sigma_1; \vec r_2,\vec\sigma_2)\nonumber \\
\quad &=& V_0 \frac{1-\vec\sigma_1\cdot\vec\sigma_2}{4} \delta^3(\vec r_1-\vec r_2)\,,
\end{eqnarray}
equal to $V_0\delta^3(\bm r)$ in case of the spin singlet ($S=0$)
and to $0$ for the spin triplet ($S=1$).

One can use the following representation of $\delta(\vec r)$
\begin{equation}
\delta^3(\vec r_1-\vec r_2)=\frac{1}{r_1r_2}
\delta(r_1-r_2)\delta(\cos\theta_1-\cos\theta_2)\delta(\phi_1-\phi_2)\,.
\end{equation}
This form is very useful when separating the angular and radial parts.
Using the relation
\begin{eqnarray}
\delta(\cos\theta_1&-&\cos\theta_2)\delta(\phi_1-\phi_2)=\nonumber \\
\nonumber \\
&=&\sum_K\frac{2K+1}{4\pi}\sum_M Y^*_{KM}(\theta_1,\phi_1) 
   Y_{KM}(\theta_2,\phi_2)\,,
\end{eqnarray}
and angular momentum algebra (see Talmi \cite{Talmi})
one gets the following energy shift of the identical ($T=1$) particles 
state $|j_1j_2J\rangle$, 
\begin{equation}
 V_\delta(J)=V_0\,F_R(n_1l_1n_2l_2)\, I(j_1j_2J)\,,
\label{eq-vdj}
\end{equation}
where
\begin{equation}
F_R(n_1l_1n_2l_2)=
\frac{1}{4\pi}\int\frac{1}{r^2}R^2_{n_1l_1}(r)R^2_{n_2l_2}(r)dr
\end{equation}
and
\begin{eqnarray}
I(j_1j_2J)=\frac{(2j_1+1)(2j_2+1)}{1+\delta_{n_1n_2}\delta_{l_1l_2}}
\left( 
\begin{array}{rrr}
   j_1 & j_2 & J\\
   1/2 & -1/2& 0
\end{array}
\right)^2\,,
\nonumber\\ 
\quad(l_1+l_2+J\,\, {\rm even})\,.
\end{eqnarray}

Now we show that the $\delta$ interaction is separable in the {\em particle-particle} 
channel.  To do this we define the creation operators of a {\it pair} of particles
coupled to the angular momentum $JM$
\begin{equation}
P^\dagger_{JM}=
\frac{1}{\sqrt{2}}\sum_{m_1m_2}(j_1m_1j_2m_2|JM)c^\dagger_{m_1} c^\dagger_{m_2}
\label{eq-pdagger}
\end{equation}
and its Hermite conjugate
\begin{equation}
P_{JM}=\left( P^\dagger_{JM} \right)^\dagger\,.
\end{equation}
Here $a^\dagger_m$ creates a nucleon in $(nljm)$ single
particle state. The $\delta$-interaction can be written in second 
quantized form 
\begin{equation}
V_{12}(\delta)=\sum_{J,\,{\rm even}} V_\delta(J) \sum_M P^\dagger_{JM}P_{JM}\,.
\label{eq-vpp}
\end{equation}
Note that $V_\delta(J)$ does not depend on the third component $M$ of 
the total angular momentum of a pair.

Since the pairing force is the interaction in the particle-particle channel 
with the third angular momentum component $M=0$, 
the operator $P^\dagger_{JM}$ defined in Eq. (\ref{eq-pdagger}) simplifies
to $P^\dagger_J\equiv P^\dagger_{J,M=0}$
and the summation in (\ref{eq-pdagger}) runs over $m=m_1=-m_2$ only.

The part of the interaction corresponding to $J=0$ is the
{\em classical pairing} or better to say monopole-pairing interaction. 
The next component $J=2$ forms the {\em quadrupole}-pairing force and so on.
The monopole pairing term reads
\begin{equation}
V_\delta(0)\sum_{m,m'}c^\dagger_m c^\dagger_{\bar m}c_{\bar m'}c_{m'}\,.
\end{equation}
The operator $c^\dagger_{\bar m}$ creates a particle in the 
time reversal state ($\bar m$) which is defined by
\begin{equation}
c_{\bar m}^\dagger=(-1)^{j-m}c_{-m}^\dagger\,.
\end{equation}
The quadrupole ($J=2$) part of Eq. (\ref{eq-vpp}) reads
\begin{equation}
V_\delta(2) \sum_{m,m'}
(j_1mj_2-m|20)(j_1m'j_2-m'|20)c^\dagger_m c^\dagger_{\bar m}c_{\bar m'}c_{m'}\,,
\end{equation}
{\em etc.} Each term in Eq. (\ref{eq-vpp}) separates in a similar 
manner. The strength $V_\delta(J)$ of each multipole is different and 
 depends on 
$V_0$ and the quantum numbers $(n_1l_1j_1)$, $(n_2l_2j_2)$ and J 
(see Eq. (\ref{eq-vdj})). 

It is easy to show that for {\it e.g.}, $j_1=j_2=9/2$
the ratio $V_\delta(2)/V_\delta(0)\approx0.24$, $V_\delta(4)/V_\delta(0)\approx0.12$
{\it etc.}. Therefore, the state $J=0$ has the largest energy shift towards the lowest 
energies (assuming $V_0$ is a positive constant).

In analogy to Eq. (\ref{eq-totham}) one can define the following operators (only even $J$ values are allowed;
see eq. (\ref{eq-vpp}))
\begin{equation}
\hat A_J = \frac{1}{2} (e^{-2i\phi} P^\dagger_{J}+e^{2i\phi} P_{J})\,, 
\qquad J=0, 2, 4, \dots\,,
\label{eq-adef}
\end{equation}
The operators $\hat A_J$ describe different multipolarities of the pairing field.
Their average values are the multipole {\it deformations} of the pairing field.
The mean pairing field Hamiltonian equivalent to the $\delta$ 
force expressed in terms of the field operators $\hat A_J$ reads
\begin{equation}
\hat V_{12}(\delta) = 
\sum_{J\,,\,\,{\rm even}} 
V_\delta(J)\alpha_J A_J  \,,
\end{equation}
where
\begin{equation}
\alpha_J=\langle\hat A_J \rangle\,
\end{equation}
is the {\em multipole deformation} of the pairing field.
The term $\hat A_0$  corresponds to the operator $\hat A$ used in the present 
paper (see Eq.(\ref{mm})) 
and describes the monopole type deformation and the parameter
$\alpha_0$ corresponds to the collective deformation parameter $\alpha$ 
used in the text
\begin{equation}
\alpha\sim\alpha_0\,.
\end{equation}

\end{document}